\newcommand{\calA}{\mathcal{A}}
\newcommand{\calC}{\mathcal{C}}
\newcommand{\calP}{\mathcal{P}}

\newcommand{\calF}{\mathcal{F}}

\newcommand{\calO}{\mathcal{O}}

\newcommand{\MI}{\mathrm{MI}}
\newcommand{\Hentropy}{\mathrm{H}}
\newcommand{\U}{\mathrm{U}}
\newcommand{\longformer}{\textsc{Longformer}}
\newcommand{\transformer}{\textsc{Transformer}}

\documentclass[11pt]{article}
\usepackage{naacl2021}
\usepackage{times}
\usepackage{here}
\usepackage{latexsym}
\usepackage{booktabs}
\usepackage{stfloats}

\usepackage{graphicx, natbib}
\usepackage{microtype}
\usepackage{amsmath}
\usepackage{amsfonts}
\usepackage{cleveref}
\usepackage{float}
\usepackage{todonotes}
\usepackage{cancel}
\usepackage{bm}
\usepackage{subfig}
\crefname{section}{\S}{\S\S}
\Crefname{section}{\S}{\S\S}
\crefname{table}{Table}{}
\crefname{figure}{Figure}{}
\crefname{algorithm}{Algorithm}{}
\crefname{equation}{eq.}{eqs.}
\crefname{appendix}{App.}{}
\crefname{prop}{Proposition}{}
\crefformat{section}{\S#2#1#3}
\usepackage{comment}
\everypar{\looseness=-1}

\newcommand{\vh}{\mathbf{h}}

\newcommand{\BERT}{\textsc{BERT}}

\everypar{\looseness=-1}

\definecolor{halsbury}{RGB}{142, 124, 193}
\definecolor{goodhart}{RGB}{194, 123, 160}

\newcommand{\citeposs}[1]{\citeauthor{#1}'s (\citeyear{#1})}

\newcommand{\saveForCR}[1]{}

\title{What About the Precedent: \\ An Information-Theoretic Analysis of Common Law}

\usepackage{tipa}

\newcommand{\ucambridge}{\normalfont \text{\textipa{D}}}
\newcommand{\ethz}{\text{\normalfont \textipa{Q}}}

\author{
Josef Valvoda$^{\ucambridge}$
~\;~ Tiago Pimentel$^{\ucambridge}$
~\;~ Niklas Stoehr$^{\ethz}$
~\;~ Ryan Cotterell$^{\ucambridge,\ethz}$ 
~\;~ Simone Teufel$^{\ucambridge}$ 
\\
\\
  $^{\ucambridge}$University of Cambridge,~\;~\;~%
  $^{\ethz}$ETH Z\"{u}rich
  \\
  \texttt{jv406@cam.ac.uk}%
  ,~\;~ \texttt{sht25@cam.ac.uk}
}

\date{}

\begin{document}
\everypar{\looseness=-1}
\setlength{\belowcaptionskip}{-4pt}
\maketitle

\begin{abstract} 
In common law, the outcome of a new case is determined mostly by precedent cases, rather than by existing statutes.
However, how exactly does the precedent influence the outcome of a new case? Answering this question is crucial for guaranteeing fair and consistent judicial decision-making. We are the first to approach this question computationally by comparing two longstanding jurisprudential views; Halsbury's, who
believes that the arguments of the precedent are the main determinant of the outcome, and Goodhart's, who believes that what matters most is the precedent's facts. 
We base our study on the corpus of legal cases from the European Court of Human Rights (ECtHR), which allows us to access not only the case itself, but also cases cited in the judges' arguments (i.e. the precedent cases). 
Taking an information-theoretic view, and modelling the question as a case outcome classification task, we find that the precedent's arguments share $0.38$ nats of information with the case's outcome, whereas precedent's facts only share $0.18$ nats of information (i.e., $58$\% less); suggesting Halsbury's view may be more accurate in this specific court. 
We found however in a qualitative analysis that there are specific statues where Goodhart's view dominates, and present some evidence these are the ones where the legal concept at hand is less straightforward.

\end{abstract}

\section{Introduction}
Legal systems around the world can be divided into two major categories \cite{matti_law}:
civil law systems, which rely predominantly on the rules written down in statutes, and common law systems, which rely predominantly on past judicial decisions, known as the precedent. 
Within common law systems, jurisprudential scholars have pondered over the nature of precedent in law for at least a century \cite{halsbury}. Is it the judges' argumentation in the precedent, or is it the claimants' specific individual circumstances that are the deciding factor in what becomes the law?
Here, we present a new information-theoretical methodology that helps answer this question.\looseness=-1

In common law countries, statutes establish the general idea of the law, but the actual scope of the law is determined by the courts during a trial. 
To keep case outcomes consistent and predictable in subsequent cases, judges are forced to apply the reasoning developed in prior cases with similar facts (precedent), to the facts of the new case under the doctrine of \emph{stare decisis} \citep{duxbury_2008, sep-legal-reas-prec, garner2009black}.
This is done by identifying the \emph{ratio decidendi} (the reasons for the decision) as opposed to the \emph{obiter dicta} (that which is
said in passing).
The distinction between ratio and obiter is an important one, since ratio is binding, whereas obiter is not. This means that courts will only strive to remain consistent in upholding ratio, but can freely depart from the obiter. 

\begin{figure}
\centering
\includegraphics[width=8.2cm]{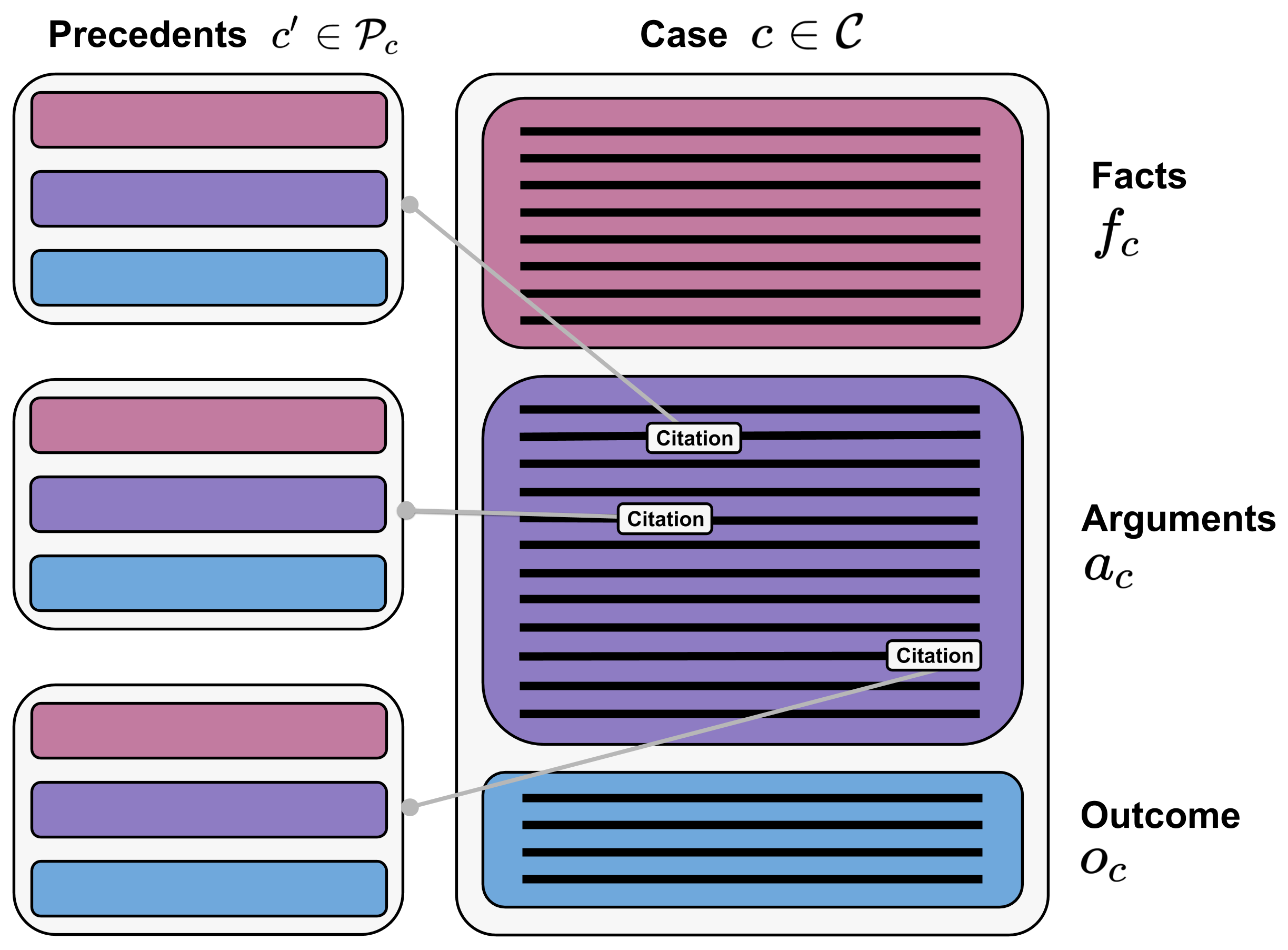}
\caption{The text of ECtHR cases can be divided into \emph{facts}, \emph{arguments} and \emph{outcome}. \emph{Arguments} cite relevant cases, also known as the \emph{precedent}.}
\label{cham}
\end{figure}

But what does the ratio consist of? There is no accepted overarching theory of precedent \citep{duxbury_2008}, but there are two tests of ratio. 
On the one hand, Lord \citet{halsbury} claims that what is binding is the judge's reasoning and arguments. For instance, by using a high degree of abstraction, judges can analogise physical and psychological pain.
A different view has been put forward by \newcite{goodhart}, who argues it is the analogy of the facts of the precedent and the case at hand, without the need for reasoning
(e.g. comparing the pain caused by a knife to that caused by another instrument, requiring a far lower degree of abstraction). 
These give rise to the two well-known legal tests for ratio: \textbf{Halsbury's test} and \textbf{Goodhart's test}. 

In this paper, we are the first to approach this problem from a data-driven perspective, using the European Court of Human Rights (\textbf{ECtHR})\footnote{European Court of Human Rights (ECtHR) is the court that adjudicates on cases dealing with the European Convention of Human Rights (ECHR).} case law; see \cref{cham}. We build a citation network over this corpus in order to have access to many precedents' full text.
Training our model on either the facts or the arguments of the precedent, we can put Halsbury's and Goodhart's views to the test.
We cast this problem as an information-theoretic study by measuring the \textbf{mutual information} \cite{shannon1948mathematical}
between the case outcome and either the precedent facts or arguments. 
We find that precedent arguments and case outcome share information to the degree of $0.38$ nats, whereas facts and case outcome only share information to the degree of $0.18$ nats (i.e., $58$\% less). We therefore observe that---at least for ECtHR---Halsbury's view of the precedent is more accurate than that of Goodhart.

\section{Legal Background}
Despite the importance of the precedent in civil law, its operationalization remains shrouded in philosophical debate centred around \emph{how} the precedent actually forms the binding law. 
Jurisprudentially, we can think of this as searching for the ratio decidendi in the judgement, i.e. separating
the ratio decidendi from the obiter dicta, or binding law from merely circumstantial statements.
It is the nature of ratio that distinguishes Halsbury's view from Goodhart's.

\subsection{Halsbury: Arguments as ratio}
The case argument contains the judge's explanation of why the case is decided the way it is. It incorporates knowledge of the precedent, facts of the case and any new reasoning the judge might develop for the case itself.
We consider the intuitive position that a legal test is formulated by the argument that the judge put forward when deciding the case.

A legal test is by its nature part of the ratio and, thus, would be binding on all subsequent cases. This is the position endorsed by Lord \citet{halsbury}. 
Under this conception of the ratio, it is the arguments that matter, becoming the law; the facts of the case are of secondary importance.
%
If a judge acts as Halsbury suggests they should extract the logic
of the implicit legal test of the precedent, and attempt to
largely ignore the specific facts of the case. 
%
Halsbury's view remains the conventional view of the precedent to this day \citep{lamond_2005}.

\subsection{Goodhart: Facts as ratio}
In contrast, \citet{goodhart} observes that many cases do not contain extensive reasoning, or any reasoning at all; judges seem to decide the outcome without these. Therefore, he claims that the facts of the case together with its outcome must form the ratio; otherwise, a hypothetical new case with the same facts as any given precedent could lead to a different outcome.
\citet{duxbury_2008} observes that judges, when in disagreement with the precedent, concentrate on the facts of a previous case more than one would expect if Halsbury's hypothesis were fully correct. Halsbury would predict that they should talk about the facts of previous cases as little as possible, and seek the most direct route to ratio in the form of argument, but they evidently do not. 
A potential explanation is that, when disagreement arises, it is easier for judges to claim that the facts are substantially different, than to challenge the logic of the precedent, i.e. to overrule that case. Overruling a previous judgement is a rare and significant legal event \cite{how_judges_overule, overruling} because it threatens the stability of the legal system. By concentrating on facts rather than running the risk of overruling, the judge can avoid this problem, including the threat of overruling her own previous judgement.

In support of this view, inspection of the argumentative part of the judgement reveals judges do not usually formulate legal tests of the kind Halsbury implies \citep{lamond_2005}. Neither do judges usually search the precedent for such legal tests \citep{alexander_sherwin_2008}.
Goodhart's position suggests that the precedent operates less as an enactment of rules, but more as reasoning by analogy; hence it is the good alignment between the facts of the two cases that leads to consistent outcomes.

\section{An Information-theoretic Approach}\label{approach}
\paragraph{Notation.}
We denote the set of cases as $\calC$, writing
each of its element as $c$. 
The set of cases that form the precedent for case $c$ are denoted $\calP_c \subset \calC$. 
We will consider three main random variables in this work.
First, we consider $O$, a random variable that ranges over a binary outcome space $\calO = \{0, 1\}^K$, where $K$ is the number of Articles. 
An instance $o \in \calO$ tells us which Articles have been violated.
Since $o$ is a vector of binary outcomes for all Articles, we can index it as $o_k$ to get the outcome of a specific $k^{\text{th}}$ Article and we analogously index the random variable $O_k$. 
We will denote $o_{c}$ the outcome of a specific case~$c$.\footnote{We note here that we overload the subscript notation in this paper. We will use subscript $c$ to denote a specific case and subscript $k$ to denote a specific article.}
Next, we consider $F$, a random variable that ranges over the space of facts. 
We denote the space of all facts as $\calF = \Sigma^*$, where $\Sigma$ is a set of sub-word units 
and $\Sigma^*$ is its Kleene closure.
We denote an instance of $F$ as $f$. 
We will further denote the facts of a specific case~$c$ as $f_c$.
Finally, we consider $A$, a random variable that ranges over the space of Arguments. 
Analogously to facts, the space of all Arguments is $\calA = \Sigma^*$.
An element of $\calA$ is denoted as $a$, which we again term~$a_c$ when referring to a specific case.

\paragraph{Operationalising Halsbury and Goodhart.}

In this work, we intend to measure the use of Halsbury's and Goodhart's views in practice, which we operationalise information-theoretically following the methodology proposed by \citet{pimentel2019meaning}.
To test the hypothesis, we construct two
collections of random variables, which we denote $H$ and $G$.
We define an instance $h_c$ of random variable $H$ as the union of arguments and outcomes for all precedent cases of $c$, i.e. $\bigcup_{c' \in \calP_c} \{ a_{c'}, o_{c'}\}$. We will denote the instance $h$ when referring to it in the abstract (without referring to a particular case). 
We analogously define instances of random variable $G$ as  
$g_c = \bigcup_{c' \in \calP_c} \{ f_{c'}, o_{c'}\}$.
While the set-theoretic notation may seem tedious, it encompasses the essence of the distinction between Halsbury's and Goodhart's view: Each view hypothesises a different group of random variables should contain more information about the outcome $O$ of a given case. 
In terms of mutual information, we are interested in comparing the following:  
\begin{align}
    \MI(O; H \mid F),\quad \MI(O; G \mid F)
\end{align}
If ${\MI(O; H \mid F) > \MI(O; G \mid F)}$, then   Halsbury's
view should be more widely used in practice.
Conversely if the opposite is true, i.e. ${\MI(O; G \mid F) > \MI(O; H \mid F)}$, then Goodhart's view should be the one more widely used.

\begin{figure}
\centering
\includegraphics[width=\columnwidth]{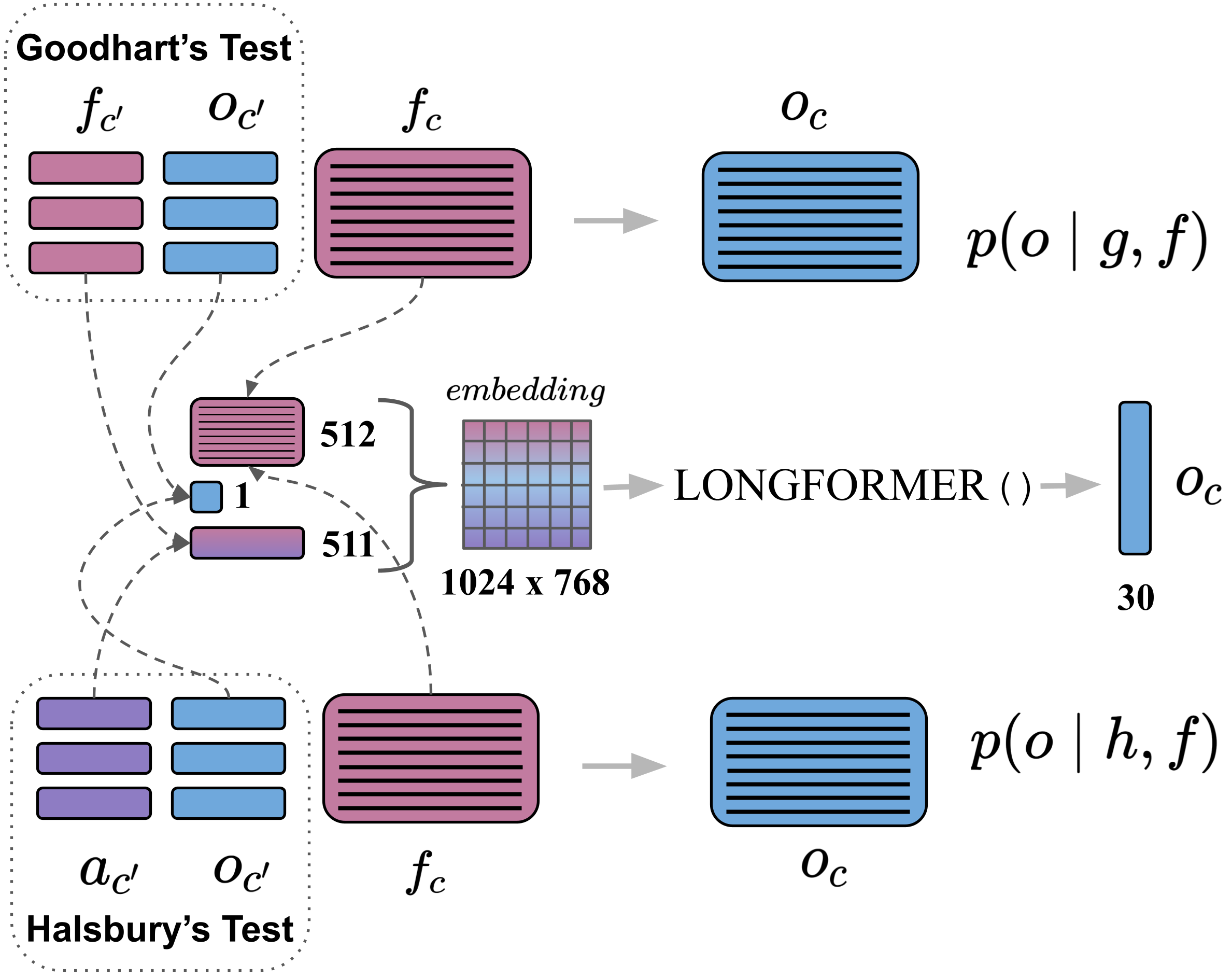}
\caption{Our formulation of Halsbury's and Goodhart's tests as a classification task. Current case facts are truncated to $512$ tokens. Outcome of the precedent is concatenated with either the precedent's facts or arguments, and both are jointly truncated at $512$ tokens. 
Finally, these are concatenated together and embedded in $768$ dimensions before being fed into the \longformer.}
\label{prediction_task}
\end{figure}

The $\MI$ is calculated by subtracting the outcome entropy conditioned on the case facts and either $H$ or $G$ from the outcome entropy conditioned on the facts alone. Therefore, to compute the $\MI$ we need to compute the Halsbury's and Goodhart's  conditional entropies first: 
\begin{align}
 \Hentropy(O \mid &\,H, F) \\
 &=-\sum_{o, h, f} p(o, h, f) \log p(o \mid h, f) \nonumber
\\[5pt]
 \Hentropy(O \mid &\,G, F)  \\
 &=-\sum_{o, g, f} p(o, g, f) \log p(o \mid g, f) \nonumber
\end{align}
as well as the entropy conditioned on the facts of the current case alone:
\begin{align}
 \Hentropy(O \mid F) &= -\sum_{o, f} p(o, f) \log p(o \mid f)
\label{entropy}
\end{align}
The conditional entropies above reflect the uncertainty (measured in nats)\footnote{Nats are computed with $\ln$, while bits use $\log_2$.} of an event, given the knowledge of another random variable. 
For instance, if $G$ completely determines $O$, then ${\Hentropy(O  \mid G)}$ is $0$; there is no uncertainty left. 
Conversely, if the variables are independent, then ${\Hentropy(O) = \Hentropy(O \mid G)}$, where $\Hentropy(O)$ denotes the unconditional entropy of the outcomes $O$.
We now note a common decomposition of mutual information that will
help with the approximation:
\begin{align}
\label{mutual_information_A}
\MI(O; H \mid F) 
&= \Hentropy(O \mid F) - \Hentropy(O \mid H, F) 
\\[5pt]
\label{mutual_information_F}
\MI(O;G \mid F)
&= \Hentropy(O \mid F) - \Hentropy(O \mid G, F)
\end{align}

In this work, we consider the conditional probabilities $p(o \mid \bullet)$ as the independent product of each Article's probability, i.e. $\prod_{k=1}^K p(o_k \mid \bullet)$. Information-theoretically, then, they are related through the following equation:
\begin{equation}
    \Hentropy(O \mid \bullet) = \sum_{k=1}^{K} \Hentropy(O_k \mid \bullet)
\end{equation}
Following \citet{williams-etal-2020-predicting}, we further calculate 
the uncertainty coefficient \cite{theil1970} of each of these mutual informations. These coefficients are easier to interpret, representing the percentage of uncertainty reduced by the knowledge of a random variable:
\begin{align}
\U(O \mid H ; F) & = \frac{\MI(O;H \mid F)}{\Hentropy(O \mid  F)}
\\[5pt]
\U(O \mid G ; F) & = \frac{\MI(O;G \mid F)}{\Hentropy(O \mid  F)}
\end{align}

\section{Experimental Setup}

We choose to work with the ECtHR corpus for three reasons. First, it can be treated as operating under precedential law, in the vein of common law countries. 
This is not a given, as the ECtHR is an international court of highest appeal without a formal doctrine of \textit{stare decisis} \cite{marc_jacob}, but there is nevertheless strong evidence that it is precedential. This evidence comes from the court's own guidelines \cite{ecthr_guide}, 
but can also be found in the writings of a former judge of the ECtHR \citep{zupancic} and of legal scholars \cite{lupu2010role}. Second, there is existing research on the neural modelling of ECtHR case law we can build upon \cite{aletras, chalkidis-etal-2019-neural, chalkidis2020legalbert}. Third, the documents of the ECtHR case law, unlike those of most other courts, textually separate the \emph{facts} from the \emph{arguments}, which is crucial for our experiments.

Case facts are descriptions of what had happened to the claimant before they went to the court; they include domestic proceedings of their case before it was appealed to the ECtHR as a form of a last resort. They do not contain any reference to European Convention of Human Rights (ECHR) Articles or ECtHR case law. Arguments on the other hand contain judges' discussion of ECHR articles and ECtHR case law in relation to the facts.
The ECtHR corpus has been scraped from the HUDOC\footnote{HUDOC: \url{https://hudoc.echr.coe.int/eng}.} database and contains $11{,}000$ cases reported in English \cite{chalkidis-etal-2019-neural}.\footnote{ECtHR cases are reported either in English, French or both. Additionally, some cases are also reported in the language of the state they take place in.} 
Judges decide for each Article of ECHR whether it has been violated with respect to the claimant's circumstances. In the ECtHR corpus, each case therefore comes with a pre-extracted decision in form of a set of violated ECHR Article numbers. We refer to this set as the \emph{outcome} of a case. Out of $30$ Articles, $18$ are from the Convention itself (Articles $2$, $3$, $4$, $5$, $6$, $7$, $8$, $9$, $10$, $11$, $12$, $13$, $14$, $18$, $34$, $38$, $41$, $46$), while the rest ($1.1$, $1.2$, $1.3$, $4.2$, $4.4$, $6.1$, $6.3$, $7.1$, $7.2$, $7.3$, $7.4$, $12.1$) comes from the Protocols  to the Convention.

For our experiment, we need a sub-corpus where each case has at least one outgoing citation where the full text is contained in our corpus. In practice, there will be other outgoing citations we cannot resolve, for instance because the document is not in English or HUDOC happens not to contain them. We also need our citations to be de-duplicated. We create such a sub-corpus, which contains $9{,}585$ documents (i.e., citing documents), with $94{,}167$ in-corpus links (tokens) to $7{,}113$ cases (types) and $65{,}495$ out-of-corpus links to $22{,}328$ types (cited documents). We start from the original ECtHR split of  $9{,}000$ training, $1{,}000$ validation and $1{,}000$ test cases, and after  citation filtering arrive at $7{,}627$ training, $976$ validation and $982$ test cases.
For every citation, we extract the text under headings with regular expressions such as  ``THE FACTS'' and ``THE LAW'', labelling it as facts and arguments, respectively. 

\subsection{Approximations}
The mutual information values that we intend to analyse need to be approximated.
We follow \citeauthor{pimentel2019meaning}'s \citeyearpar{pimentel2019meaning,pimentel-etal-2021-finding} methodology for this, approximating them as the difference between two cross-entropies:
\begin{align}
\MI(O;H \mid F) 
&\phantom{:}\approx \Hentropy_{\theta}(O \mid F) - \Hentropy_{\theta}(O \mid H, F) \nonumber
\\[5pt]
\MI(O;G \mid F)
&\phantom{:}\approx \Hentropy_{\theta}(O \mid F) - \Hentropy_{\theta}(O \mid G, F) \nonumber
\end{align}
Indeed, although several estimates for the mutual information exist, \citet{mcallester2020formal} argues that estimating it as this difference is the most  statistically justified way. These conditional entropies are themselves approximated through their sample estimate. For instance, we compute:
\begin{align}
\Hentropy_{\theta}(O  \mid G, F) &\approx -\frac{1}{|\mathcal{C}|}\sum_{c \in \calC} \log p_\theta (o_c \mid g_c, f_c)
\end{align}
which is exact as $|\mathcal{C}|\rightarrow\infty$.
We note that the cross-entropy is an upper bound on the entropy, which uses a model $p_\theta (o \mid \bullet)$ for its estimate. The better this model, the tighter our estimates will be.
The only thing left to do now, is to obtain these probability estimates.
We thus model Halsbury's view as a classification task (see \cref{prediction_task}) estimating the probability:
\begin{align}
    \label{goodhart_test}
    p_\theta(o \mid h, f)
    &=  \prod_{k=1}^{K}  p_\theta(o_{k} \mid h, f)
\end{align}
We analogously model Goodhart's view as:
\begin{align}
    \label{halsbury_test}
    p_\theta(o \mid g, f)
    &=  \prod_{k=1}^{K}  p_\theta(o_{k} \mid g, f)
\end{align}
Finally, we model the $p_{\theta}$ of the model conditioned only on the facts of the case at hand as:
\begin{align}
    \label{baseline_test}
    p_\theta(o \mid f)
    &=  \prod_{k=1}^{K}  p_\theta(o_{k} \mid f)
\end{align}
These models can be approximated using deep neural networks as introduced in the next section. We train deep neural networks on our training sets, using a cross-entropy loss function and a sub-gradient descent method. Given the trained models, we can then answer if it is Halsbury's view or Goodhart's that is more widely used by the ECtHR judiciary.

\subsection{Implementation Details}


All experiments are conducted using a \longformer\phantom{a}classifier \cite{longformer}.\footnote{Our code is available here: \url{https://github.com/valvoda/Precedent.}} The \longformer\phantom{a}is built on the same \transformer\phantom{a}\cite{vaswani2017attention} architecture as \BERT\phantom{a}\cite{devlin-etal-2019-bert}, but allows for up to $4{,}096$ tokens, using an attention mechanism which scales linearly, instead of quadratically. 
We choose this architecture in particular as it achieves state-of-the-art performance in tasks similar to ours, e.g. on the IMDB sentiment classification \cite{maas-etal-2011-learning} and Hyperpartisan news detection \cite{kiesel-etal-2019-semeval}.

To find the probability of violation of the $K$ Articles we compute:
\begin{align} \label{eq:lognformer}
    &\vh = \longformer(g, f) \\
    &p_\theta(o \mid g, f) = \sigma(W^{(1)}\,\mathrm{ReLU}(W^{(2)}\,\vh)) \nonumber
\end{align}
where $\vh \in \mathbb{R}^{d_1}$ is a high dimensional representation, $W^{(1)} \in \mathbb{R}^{K \times d_2}$ and $W^{(2)} \in \mathbb{R}^{d_2 \times d_1}$
are learnable parameters in linear projections, and $\sigma$ is the sigmoid function.
\Cref{eq:lognformer} will thus output a $K$-dimensional vector with the probabilities for all articles, by indexing this vector we retrieve the probabilities of the individual articles applying.
Due to resource limitations we set the models' hidden size to $50$ and batch size to $16$, and also truncate individual cases to $512$ tokens. 
%
For the models $p_\theta(o_k \mid g, f)$ and $p_\theta(o_k \mid h, f)$, which are trained on the combination of $f$ and either $h$ or $g$, we concatenate cases to the maximum length of $1{,}024$ tokens (as exemplified in \cref{prediction_task}).
While we do not fully utilise the $4{,}096$ word limit of the \longformer, we are able to process twice as many tokens as standard \BERT\phantom{a}without pooling; memory limitations prevent us from using the full $4{,}096$ tokens, though.

Our \longformer\phantom{a}models are implemented using the Pytorch \cite{paszke2019pytorch} and Hugginface \cite{Wolf2019HuggingFacesTS} Python libraries. We train all our models on $4$ Nvidia P$100$ $16$GiB GPU's for a maximum of $6$ hours using \longformer-base model. Our results are reported in terms of the models cross entropy.
\begin{table}[ht]
  \centering
  \begin{tabular}{p{2.2cm} p{1.1cm} p{1.1cm} p{1.1cm}}
  \toprule
    \textbf{Model Input} & $\Hentropy_\theta$ & $\MI$ & $\U$ \\
  \midrule
    \textbf{Facts} & 2.99 & - & - \\

    \textbf{\textcolor{goodhart}{Goodhart}} & 2.81 & 0.18 & 6\% \\

    \textbf{\textcolor{halsbury}{Halsbury}} & 2.68 & 0.31 & 10\% \\
    
    \bottomrule
  \end{tabular}
  \caption{The cross entropy $\Hentropy_\theta$, mutual information $\MI$ and uncertainty coefficient $\U$ results.}
  \label{tab:2}
\end{table}

\section{Results}
Our experimental results are contained in \cref{tab:2}. 
We first note that both our mutual information estimates are statistically larger than zero, i.e. Goodhart's and Halsbury's cross-entropies are statistically smaller than that of the Facts.\footnote{We measure significance using the two tailed paired permutation tests with $p < 0.05$ after \citeposs{benjamini1995controlling} correction.} 
The question we asked ourselves at the outset, though, concerns whether the data supports Halsbury's or Goodhart's view. 
We find that our estimate of $\MI(O;H \mid F)$ is significantly
larger at $0.31$ nats than our estimate of  $\MI(O;G \mid F)$ at $0.18$ nats.
These results suggest that the information contributed by the precedent arguments give us nearly $58$\% more information about the outcome of the case than the information contained in the facts of the precedent. In terms of the uncertainty coefficient, the outcome entropy is reduced by 6\% for facts and by 10\% for arguments. We therefore observe that Halsbury's view is more widely used in the domain of ECtHR than Goodhart's.

\section{Discussion \& Analysis}
\begin{figure}[tp]
    \centering
    \includegraphics[width=7.5cm]{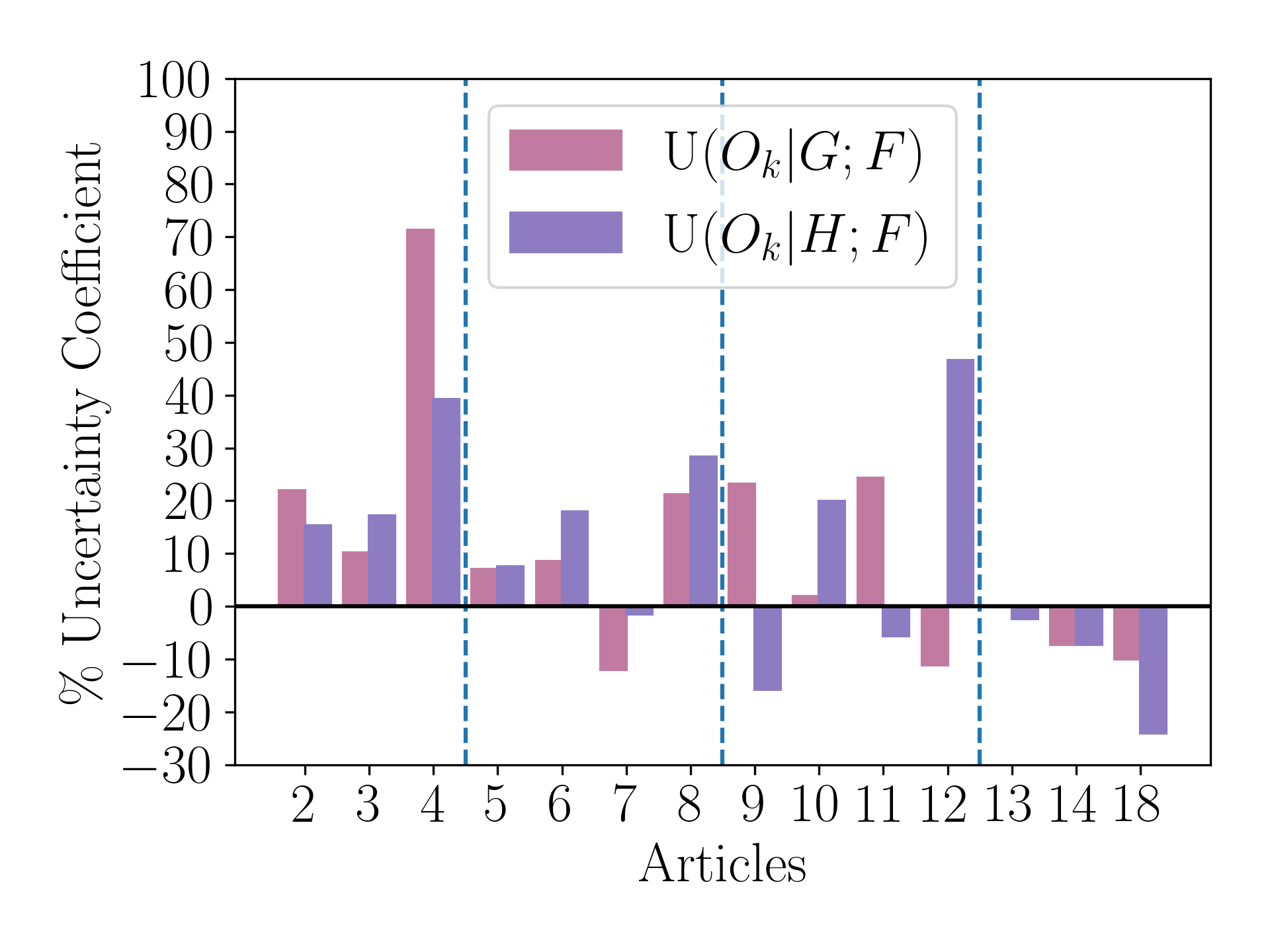}
    \caption{Uncertainty coefficient for the Articles of the ECHR Convention.}
    \label{fig:bar_all}
\end{figure}

A more nuanced story can be told if we inspect the individual Articles even though the small number of cases per Article does not allow for conclusive significance tests.
The core rights of the Convention are contained in (Articles $2$-$18$).\footnote{The Convention Section $1$ is the first section of the ECHR and elevates some of the Universal Declaration of Human Rights principles into actionable rights of European citizens \cite{schindler1962european}.}
\cref{fig:bar_all} shows that for some of the core Articles, we see the opposite effect from what we observed for the entirety of Articles, namely that facts outperform arguments, in particular for Articles $2$, $4$, $9$, $11$, $13$, $18$.\looseness=-1

We hypothesise the reason for this is either that the judges have not yet developed a functional legal method for these Articles, that the relevant precedent has been placed late in the list of precedents (and thus was truncated away by our methodology), or that the complexity of the arguments requires a reasoning ability our models are simply not capable of. We consider each hypothesis separately below.

\subsection{Conceptual Uncertainty}
For some Articles, it is more difficult to develop a legal method than for others because the logic of the argument is elusive for some reason. This holds, for instance, for Articles encoding a vague concept such as ``right to life'', cf. the discussion below. If a case deals with such an Article, the argument of a potential precedent will be less useful to determine the outcome. We hypothesise that in such a case the judges will be more willing to depart from the logic of past cases, which they might perceive as unsatisfactory in search of a better legal reasoning.
However, judges strive to maintain consistency between decisions as their authority is based on this consistency. 
Under these conditions, a judge might take the approach of trying to find precedent cases that match the current case in terms of facts even if not in terms of logic. 
Case law dealing with such Articles would therefore be more likely to follow Goodhart's view.

To support or disprove this hypothesis would require an in-depth legal analysis far beyond the scope of this paper; one would need to robustly argue why judges find it relatively more difficult to develop legal reasoning for certain articles. 
However, looking at the Articles where our data indicate that Goodhart's view is the one more widely used, it seems to us that they indeed concern legal concepts that are more slippery than others, which we categorised as follows.

\subsubsection{Corporal Articles}
We can contrast Articles $2$ and $4$, where judges follow Goodhart's view, to Article $3$, for which judges follow Halsbury's view instead, see \cref{tab:3}. 
All three Articles are concerned with the fundamental respect of human life, and we therefore consider them together as the \emph{corporal} Articles.

Article $2$: Right to Life prohibits the \emph{intentional deprivation of life, save for circumstances where it is a penalty for a crime, in defence, during an arrest, or riot suppression}. In the context of the criminal code of Europe, this is a very restricted prohibition. Every country already encodes these rules. On the other hand, it raises the difficult issues of beginning and end of life. Is Article $2$ for or against abortion \cite{cosentino2015safe}? What is its stance on euthanasia \cite{euthanasia}? Developing a legal test for Article $2$ seems very hard indeed.\looseness=-1

Similarly, Article $4$: \emph{Prohibition of slavery and forced labour}, excludes work forced in detention, compulsory military service, any service during emergency or ``normal'' civic obligations. Due to the large number of exceptions to the general rule it seems very hard to establish what exactly this Article does prohibit.

Let us compare these to Article $3$: \emph{Prohibition of torture}, where Halsbury's view prevails. This Article simply states that \emph{no one shall be subjected to torture or inhuman or degrading treatment or punishment}. No exceptions are given. It seems much easier to develop a legal test for Article $3$ than for Articles $2$ and $4$. The judges are free to establish what constitutes torture; whereas when it comes to Articles $2$ and $4$, they are facing many restrictions---both legal and political. 

\subsubsection{Faith and Family Articles}
Above, we compare Articles concerned with corporal matters. In a similar way we can also group Articles $8$, $9$, $10$, $11$ and $12$ as the Articles broadly concerning belief, family and religion. 

The two outliers here are Articles $9$ and $11$. Article $9$ provides the \emph{freedom of thought, conscience and religion}, Article $11$ provides the \emph{freedom of assembly and association}. 
For both Articles, Goodhart's test outperforms Halsbury's.

Just like above, the nature of Articles $9$ and $11$ seems more complicated compared to Article $8$, which is  similar, but narrower in scope: \emph{Right to respect for private and family life}, Article $10$: \emph{Freedom of expression} and Article $12$: \emph{Right to marry}. 

We would argue that since Articles $8$ and $12$ provide a \emph{right} as opposed to a \emph{freedom}, they define more narrowly the obligation on the part of the State. Compared to the \emph{freedom of thought} and \emph{association} (Articles $9$ and $11$), the \emph{right to marry} and the right to \emph{privacy} (Article $8$ and $12$) seem to be more concrete and testable obligations.

We can further view Article $10$: \emph{freedom of expression}, as dealing with an action brought about by the exercise of  Article $9$: \emph{freedom of thought}. While similar in concept, regulating speech seems far easier in practice than regulating thought.

Finally, an inspection of the ECHR guidelines to Article $11$ reveals that judges seem to be often torn between Articles $10$ and $11$.\footnote{Article $11$ guidance: \url{https://www.echr.coe.int/Documents/Guide_Art_11_ENG.pdf}.} This is because much of the cases dealing with Article $11$ concern themselves with disentangling what constitutes an expression during an assembly and conversely which assembly is a form of an expression. Many cases deal with the question of religious gathering as an assembly. This is obviously not an easy position for a judge to divine a legal test for, and perhaps a good reason for turning to the facts of the precedent cases for consistency instead.

\begin{table}[tp]
  \centering
  \resizebox{\columnwidth}{!}{%
  \begin{tabular}{r c r r r r}
  \toprule
   & & \multicolumn{2}{c}{\textbf{\textcolor{goodhart}{Goodhart}}} & \multicolumn{2}{c}{\textbf{\textcolor{halsbury}{Halsbury}}} \\ \cmidrule(lr{.5em}){3-4} \cmidrule(lr{.5em}){5-6}
    \textbf{Art} & $\Hentropy_\theta(O_k \mid F)$  & \multicolumn{1}{c}{$\MI$} & \multicolumn{1}{c}{$\U$} & \multicolumn{1}{c}{$\MI$} & \multicolumn{1}{c}{$\U$} \\
    \midrule
\textbf{2} & 0.065 & 0.014 & 21.97\% & 0.010 & 15.27\% \\ 
\textbf{3} & 0.272 & 0.028 & 10.15\% & 0.047 & 17.23\% \\ 
\textbf{4} & 0.028 & 0.020 & 71.26\% & 0.011 & 39.27\% \\ 
 \midrule
\textbf{5} & 0.275 & 0.019 & 7.05\% & 0.021 & 7.53\% \\ 
\textbf{6} & 0.493 & 0.042 & 8.50\% & 0.089 & 17.95\% \\ 
\textbf{7} & 0.024 & -0.003 & -12.01\% & -0.000 & -1.52\% \\ 
 \midrule
\textbf{8} & 0.298 & 0.063 & 21.15\% & 0.084 & 28.33\% \\ 
\textbf{9} & 0.022 & 0.005 & 23.14\% & -0.003 & -15.74\% \\ 
\textbf{10} & 0.173 & 0.003 & 1.92\% & 0.034 & 19.90\% \\ 
\textbf{11} & 0.074 & 0.018 & 24.29\% & -0.004 & -5.66\% \\ 
\textbf{12} & 0.006 & -0.001 & -11.09\% & 0.003 & 46.60\% \\ 
 \midrule
\textbf{13} & 0.235 & -0.000 & -0.10\% & -0.006 & -2.38\% \\ 
\textbf{14} & 0.071 & -0.005 & -7.30\% & -0.005 & -7.28\% \\ 
\textbf{18} & 0.031 & -0.003 & -10.00\% & -0.007 & -24.01\% \\ 
\bottomrule
  \end{tabular}
  }
  \caption{The cross-entropy $\Hentropy_\theta$, mutual information $\MI$ and uncertainty coefficient $\U$ results of each of the core ECHR Articles. We note that these values are empirical estimates, so negative $\MI$ results are caused by an approximation error in our models.
  }
  \label{tab:3}
\end{table}

\subsection{Late Precedent}
There is a group of Articles in the last quarter of \cref{fig:bar_all} ($13$, $14$, $18$) for which neither Goodhart's nor Halsbury's view seem to hold. We speculate that the reason for this is that these Articles never appear alone, and instead always appear in 
conjunction with another Article, and also that they appear late in the list of precedents, so get truncated with our methodology. 

Articles $13$: \emph{Right to an effective remedy}, $14$: \emph{Prohibition of discrimination} and $18$: \emph{Limitation on use of restrictions on rights}, are designed to ensure that states provide remedy for their wrongdoing, equal access to the rights, and do not use the restrictions in Articles for Human Rights abuse. 

To claim any one of these Articles, the claimant will also have to claim a violation of one of the primary Articles as their core grievance for which they seek the remedy or equal treatment, for instance Article $3$: \emph{Prohibition of torture}. This means that any case dealing with Articles $13$, $14$ and $18$ is likely to focus on the violation of that primary right.

While there might be a precedent present for the secondary Articles, the probability is high that our models will not have the chance to train on them because they appear late and because our method truncates text due to computational complexity reasons. This could explain why for these Articles, all our models trained on the precedent cases underperform when compared to the models trained on the facts of the case alone.

\subsection{Model Limitations}
Another possible explanation for the different behaviour between Articles could lie within the limitations of the neural architecture. There could be a model bias for facts in precedent since they are more similar to the facts at hand as opposed to the arguments. If this is the case our results understate the value of arguments. While this is a concern, the overall results of our paper would not change even if we could remove this bias since we find arguments more important than facts despite this potential handicap.

On a more nuanced level, Articles $2$ and~$4$ above might require a higher level of reasoning than their Article~$3$ counterpart. 
So while the judges might have developed a satisfying legal test for them, our models simply aren't able to learn it. For example for Article $7$: \emph{No punishment without law}, our precedent models fail to learn any additional information from the precedent facts or arguments.

This might simply be the result of an insufficient representation of Article $7$ in training cases, or of its appearance truncated out of the input. However it also raises the question of what a \transformer\phantom{a}model can learn. 

The nascent field of \emph{BERTology} has explored exactly this question \cite{rogers2020primer, pimentel-etal-2020-information}.
In particular the work of \citet{niven-kao-2019-probing}, examining \BERT\phantom{a}performance on the English Argument Reasoning Comprehension Task \cite{habernal-etal-2018-argument}, suggest that instead of \BERT\phantom{a}being able to reason, it is merely very good at utilising the artefacts in the data when compared to  previous approaches. As \citet{bender-koller-2020-climbing} contend a system can't ever learn meaning from form alone.
According to their view, description of the case facts alone will never fully capture the reality of the world the claimant inhabits.

On the other hand, there is some evidence towards transformers being able to reason over simple sentences \citet{clark_transform}. While this is encouraging, legal documents are far more complicated than the simple sentences considered in the study above. Either way, the models' ability to reason in the way a human lawyer would is certainly limited and could explain the diminished performance for the more complicated Articles.

\section{Related work}

In this section, we contextualise our work with relation to the related research on legal AI.
Computational approaches to solving legal problems go back at least as far as the late 1950's \citep{kort_1957, nagel1963applying}. 
Early research has focused on crafting rule-based systems for case outcome prediction, achieving human-like performance by the early 2000's \citep{ashley}. These systems however proved too brittle to keep up with the ever-changing legal landscape and never transitioned from research into industry.

More recently, a new wave of deep learning methods has reinvigorated the research interest in legal AI. The majority of this new work has been conducted on statutory legal systems which do not rely on the doctrine of precedent to nearly the same extent as their common law counterparts. For instance, in Chinese law the use of neural models for case outcome classification has already been investigated extensively \cite{hu-etal-2018-shot, zhong-etal-2018-legal, xu2020distinguish}.
In the precedential legal domain, smaller corpora of annotated cases have been investigated over the years \citep{grover, Valvoda18}. 
However, large-scale corpora necessary for deep learning architectures have become available only recently. The Caselaw Access Project\footnote{Caselaw Access Project:, \url{https://case.law}} introduced a large dataset of American case law in 2018. \citet{aletras} have introduced the ECtHR corpus, and \citet{chalkidis-etal-2019-neural} have run deep neural networks on it in order to predict outcome. Similarly, the Canadian Supreme Court Case corpus has been used in information retrieval for the first time by \citet{coliee}. This improved access to a high quality common law datasets has opened up a potential for new work in the field of legal AI.

Particularly similar to our work is the study done by \citet{sim-etal-2016-friends}, who have considered the influence of
petitioners and responders (amicus) briefs on the US Supreme Court decision and opinions.

\section{Conclusion}

In this paper, we have shifted the focus of legal AI research from practical tasks such as precedent retrieval or outcome prediction, to a theoretical question: which aspect of the precedent is most important in forming the law? To this end, we trained a similar neural modelling approach as \citet{chalkidis-etal-2019-neural} to predict the outcome of a case on the ECtHR dataset, and inspected the difference in the mutual information between our operationalisations of Halsbury's and Goodhart's view. We have used a method inspired by \citet{pimentel2019meaning} to approximate the $\MI$. 
We observe that out of the two archetypal views on precedent, that of Halsbury and Goodhart, the former has a better empirical support in the domain of ECtHR case law.\looseness=-1

This study has demonstrated a novel method of approaching jurisprudential questions using the information-theoretic toolkit. 
We hope that future work can leverage our methodology towards answering other questions of legal philosophy. However, our results are not only of an interest in the context of legal theory, but they can also inform a development of better legal models in practice. 
Since most precedential reasoning is conducted using the arguments in the precedent, outcome prediction models should take advantage of the case arguments, instead of relying solely on the facts.



\section*{Ethical Considerations}
While our work is not concerned with a legal application, it is important to note that the results presented here are qualified by the limitations of contemporary NLP models' ability to process language. It should therefore serve as no indication that judges could (or should) be replaced by models or techniques discussed in this paper.

\section*{Acknowledgements}
We are grateful to Prof. Ken Satoh for all the fruitful discussions leading towards this paper. We further thank the National Institute of Informatics (NII) Japan and Huawei research UK for their financial support enabling this research.

\bibliographystyle{acl_natbib}
\bibliography{thesis.bib}

\begin{thebibliography}{50}
\expandafter\ifx\csname natexlab\endcsname\relax\def\natexlab#1{#1}\fi

\bibitem[{Aletras et~al.(2016)Aletras, Tsarapatsanis, Preo{\c t}iuc-Pietro, and
  Lampos}]{aletras}
Nikolaos Aletras, Dimitrios Tsarapatsanis, Daniel Preo{\c t}iuc-Pietro, and
  Vasileios Lampos. 2016.
\newblock \href {https://doi.org/10.7717/peerj-cs.93} {{Predicting judicial
  decisions of the European Court of Human Rights: a Natural Language
  Processing perspective}}.
\newblock \emph{PeerJ Computer Science}, 2:e93.

\bibitem[{Alexander and Sherwin(2008)}]{alexander_sherwin_2008}
Larry Alexander and Emily Sherwin. 2008.
\newblock \href {https://doi.org/10.1017/CBO9781139167420.005} {\emph{The
  Mystification of Common-Law Reasoning}}, Cambridge Introductions to
  Philosophy and Law, page 64–103. Cambridge University Press.

\bibitem[{Ashley(2017)}]{ashley}
Kevin~D. Ashley. 2017.
\newblock \href {https://doi.org/10.1017/9781316761380} {\emph{Artificial
  Intelligence and Legal Analytics}}.
\newblock Cambridge University Press, Cambridge.

\bibitem[{Beltagy et~al.(2020)Beltagy, Peters, and Cohan}]{longformer}
Iz~Beltagy, Matthew~E. Peters, and Arman Cohan. 2020.
\newblock \href {http://arxiv.org/abs/2004.05150} {{Longformer: The
  Long-Document Transformer}}.
\newblock \emph{arXiv}.

\bibitem[{Bender and Koller(2020)}]{bender-koller-2020-climbing}
Emily~M. Bender and Alexander Koller. 2020.
\newblock \href {https://doi.org/10.18653/v1/2020.acl-main.463} {Climbing
  towards {NLU}: {On} meaning, form, and understanding in the age of data}.
\newblock In \emph{Proceedings of the 58th Annual Meeting of the Association
  for Computational Linguistics}, pages 5185--5198, Online. Association for
  Computational Linguistics.

\bibitem[{Benjamini and Hochberg(1995)}]{benjamini1995controlling}
Yoav Benjamini and Yosef Hochberg. 1995.
\newblock \href
  {https://www.jstor.org/stable/2346101?seq=1#metadata_info_tab_contents}
  {Controlling the false discovery rate: a practical and powerful approach to
  multiple testing}.
\newblock \emph{Journal of the Royal Statistical Society: Series B
  (Methodological)}, 57(1):289--300.

\bibitem[{Black(2019)}]{garner2009black}
Henry Black. 2019.
\newblock \href {https://thelawdictionary.org/} {\emph{Black's Law
  Dictionary}}, 11th edition.
\newblock Thomson Reuters.

\bibitem[{Chalkidis et~al.(2019)Chalkidis, Androutsopoulos, and
  Aletras}]{chalkidis-etal-2019-neural}
Ilias Chalkidis, Ion Androutsopoulos, and Nikolaos Aletras. 2019.
\newblock \href {https://doi.org/10.18653/v1/P19-1424} {Neural legal judgment
  prediction in {E}nglish}.
\newblock In \emph{Proceedings of the 57th Annual Meeting of the Association
  for Computational Linguistics}, pages 4317--4323, Florence, Italy.
  Association for Computational Linguistics.

\bibitem[{Chalkidis et~al.(2020)Chalkidis, Fergadiotis, Malakasiotis, Aletras,
  and Androutsopoulos}]{chalkidis2020legalbert}
Ilias Chalkidis, Manos Fergadiotis, Prodromos Malakasiotis, Nikolaos Aletras,
  and Ion Androutsopoulos. 2020.
\newblock \href {https://doi.org/10.18653/v1/2020.findings-emnlp.261}
  {{LEGAL}-{BERT}: The muppets straight out of law school}.
\newblock In \emph{Findings of the Association for Computational Linguistics:
  EMNLP 2020}, pages 2898--2904, Online. Association for Computational
  Linguistics.

\bibitem[{Clark et~al.(2020)Clark, Tafjord, and Richardson}]{clark_transform}
Peter Clark, Oyvind Tafjord, and Kyle Richardson. 2020.
\newblock \href {https://doi.org/10.24963/ijcai.2020/537} {{Transformers as
  Soft Reasoners over Language}}.
\newblock In \emph{Proceedings of the Twenty-Ninth International Joint
  Conference on Artificial Intelligence, {IJCAI-20}}, pages 3882--3890.
  International Joint Conferences on Artificial Intelligence Organization.
\newblock Main track.

\bibitem[{Cosentino(2015)}]{cosentino2015safe}
Chiara Cosentino. 2015.
\newblock \href {https://academic.oup.com/hrlr/article/15/3/569/2356117} {Safe
  and legal abortion: An emerging human right? the long-lasting dispute with
  state sovereignty in {ECHR} jurisprudence}.
\newblock \emph{Human Rights Law Review}, 15(3):569--589.

\bibitem[{Devlin et~al.(2019)Devlin, Chang, Lee, and
  Toutanova}]{devlin-etal-2019-bert}
Jacob Devlin, Ming-Wei Chang, Kenton Lee, and Kristina Toutanova. 2019.
\newblock \href {https://doi.org/10.18653/v1/N19-1423} {{BERT}: Pre-training of
  deep bidirectional transformers for language understanding}.
\newblock In \emph{Proceedings of the 2019 Conference of the North {A}merican
  Chapter of the Association for Computational Linguistics: Human Language
  Technologies, Volume 1 (Long and Short Papers)}, pages 4171--4186,
  Minneapolis, Minnesota. Association for Computational Linguistics.

\bibitem[{Dunn(2003)}]{how_judges_overule}
Pintip~Hompluem Dunn. 2003.
\newblock \href {http://www.jstor.org/stable/3657527} {How judges overrule:
  Speech act theory and the doctrine of stare decisis}.
\newblock \emph{The Yale Law Journal}, 113(2):493--531.

\bibitem[{Duxbury(2008)}]{duxbury_2008}
Neil Duxbury. 2008.
\newblock \href {https://doi.org/10.1017/CBO9780511818684.005}
  {\emph{Distinguishing, overruling and the problem of self-reference}}, page
  111–149. Cambridge University Press.

\bibitem[{ECtHR(2014)}]{ecthr_guide}
ECtHR. 2014.
\newblock \href
  {https://www.echr.coe.int/Documents/Guide\_ECHR\_lawyers\_ENG.pdf} {European
  court of human rights: Questions \& answers for lawyers}.

\bibitem[{Goodhart(1930)}]{goodhart}
Arthur~L. Goodhart. 1930.
\newblock \href {http://www.jstor.org/stable/790205} {Determining the ratio
  decidendi of a case}.
\newblock \emph{The Yale Law Journal}, 40(2):161--183.

\bibitem[{Grover et~al.(2003)Grover, Hachey, Hughson, and Korycinski}]{grover}
Claire Grover, Ben Hachey, Ian Hughson, and Chris Korycinski. 2003.
\newblock \href {https://doi.org/10.1145/1047788.1047839} {Automatic
  summarisation of legal documents}.
\newblock In \emph{Proceedings of the 9th International Conference on
  Artificial Intelligence and Law}, ICAIL '03, page 243–251, New York, NY,
  USA. Association for Computing Machinery.

\bibitem[{Habernal et~al.(2018)Habernal, Wachsmuth, Gurevych, and
  Stein}]{habernal-etal-2018-argument}
Ivan Habernal, Henning Wachsmuth, Iryna Gurevych, and Benno Stein. 2018.
\newblock \href {https://doi.org/10.18653/v1/N18-1175} {The argument reasoning
  comprehension task: Identification and reconstruction of implicit warrants}.
\newblock In \emph{Proceedings of the 2018 Conference of the North {A}merican
  Chapter of the Association for Computational Linguistics: Human Language
  Technologies, Volume 1 (Long Papers)}, pages 1930--1940, New Orleans,
  Louisiana. Association for Computational Linguistics.

\bibitem[{Halsbury(1907)}]{halsbury}
Lord Halsbury. 1907.
\newblock \emph{Halsbury's Laws of England}, 1 edition.
\newblock LexisNexis Butterworths.

\bibitem[{Hendriks(2019)}]{euthanasia}
Arend~Cornelis Hendriks. 2019.
\newblock \href {https://doi.org/10.1007/s12027-018-0530-7} {{End-of-life
  decisions. Recent jurisprudence of the European Court of Human Rights}}.
\newblock \emph{ERA Forum}, 19(4):561--570.

\bibitem[{Hu et~al.(2018)Hu, Li, Tu, Liu, and Sun}]{hu-etal-2018-shot}
Zikun Hu, Xiang Li, Cunchao Tu, Zhiyuan Liu, and Maosong Sun. 2018.
\newblock \href {https://www.aclweb.org/anthology/C18-1041} {Few-shot charge
  prediction with discriminative legal attributes}.
\newblock In \emph{Proceedings of the 27th International Conference on
  Computational Linguistics}, pages 487--498, Santa Fe, New Mexico, USA.
  Association for Computational Linguistics.

\bibitem[{Jacob(2014)}]{marc_jacob}
Marc Jacob. 2014.
\newblock \href {https://doi.org/10.1017/CBO9781107053762} {\emph{Precedents
  and Case-Based Reasoning in the European Court of Justice: Unfinished
  Business}}.
\newblock Cambridge University Press.

\bibitem[{Joutsen(2019)}]{matti_law}
Matti Joutsen. 2019.
\newblock \href {https://doi.org/10.1017/9781108597296} {\emph{International
  and Transnational Crime and Justice}}, 2 edition.
\newblock Cambridge University Press.

\bibitem[{Kiesel et~al.(2019)Kiesel, Mestre, Shukla, Vincent, Adineh, Corney,
  Stein, and Potthast}]{kiesel-etal-2019-semeval}
Johannes Kiesel, Maria Mestre, Rishabh Shukla, Emmanuel Vincent, Payam Adineh,
  David Corney, Benno Stein, and Martin Potthast. 2019.
\newblock \href {https://doi.org/10.18653/v1/S19-2145} {{S}em{E}val-2019 task
  4: Hyperpartisan news detection}.
\newblock In \emph{Proceedings of the 13th International Workshop on Semantic
  Evaluation}, pages 829--839, Minneapolis, Minnesota, USA. Association for
  Computational Linguistics.

\bibitem[{Kort(1957)}]{kort_1957}
Fred Kort. 1957.
\newblock \href {https://doi.org/10.2307/1951767} {Predicting supreme court
  decisions mathematically: A quantitative analysis of the “right to
  counsel” cases}.
\newblock \emph{American Political Science Review}, 51(1):1–12.

\bibitem[{Lamond(2005)}]{lamond_2005}
Grant Lamond. 2005.
\newblock \href {https://doi.org/10.1017/S1352325205050019} {Do precedents
  create rules?}
\newblock \emph{Legal Theory}, 11(1):1–26.

\bibitem[{Lamond(2016)}]{sep-legal-reas-prec}
Grant Lamond. 2016.
\newblock \href
  {https://stanford.library.sydney.edu.au/entries/legal-reas-prec/} {{Precedent
  and Analogy in Legal Reasoning}}.
\newblock In Edward~N. Zalta, editor, \emph{The {Stanford} Encyclopedia of
  Philosophy}, spring 2016 edition. Metaphysics Research Lab, Stanford
  University.

\bibitem[{Lupu and Voeten(2010)}]{lupu2010role}
Yonatan Lupu and Erik Voeten. 2010.
\newblock \href {https://opensiuc.lib.siu.edu/pnconfs_2010/12/} {The role of
  precedent at the european court of human rights: A network analysis of case
  citations}.

\bibitem[{Maas et~al.(2011)Maas, Daly, Pham, Huang, Ng, and
  Potts}]{maas-etal-2011-learning}
Andrew~L. Maas, Raymond~E. Daly, Peter~T. Pham, Dan Huang, Andrew~Y. Ng, and
  Christopher Potts. 2011.
\newblock \href {https://www.aclweb.org/anthology/P11-1015} {Learning word
  vectors for sentiment analysis}.
\newblock In \emph{Proceedings of the 49th Annual Meeting of the Association
  for Computational Linguistics: Human Language Technologies}, pages 142--150,
  Portland, Oregon, USA. Association for Computational Linguistics.

\bibitem[{McAllester and Stratos(2020)}]{mcallester2020formal}
David McAllester and Karl Stratos. 2020.
\newblock \href {https://arxiv.org/abs/1811.04251} {Formal limitations on the
  measurement of mutual information}.
\newblock In \emph{International Conference on Artificial Intelligence and
  Statistics}, pages 875--884.

\bibitem[{Nagel(1963)}]{nagel1963applying}
Stuart~S. Nagel. 1963.
\newblock \href
  {https://heinonline.org/HOL/Page?handle=hein.journals/tlr42&div=63&g_sent=1&casa_token=&collection=journals}
  {Applying correlation analysis to case prediction}.
\newblock \emph{Texas Law Review}, 42:1006.

\bibitem[{Niven and Kao(2019)}]{niven-kao-2019-probing}
Timothy Niven and Hung-Yu Kao. 2019.
\newblock \href {https://doi.org/10.18653/v1/P19-1459} {Probing neural network
  comprehension of natural language arguments}.
\newblock In \emph{Proceedings of the 57th Annual Meeting of the Association
  for Computational Linguistics}, pages 4658--4664, Florence, Italy.
  Association for Computational Linguistics.

\bibitem[{Paszke et~al.(2019)Paszke, Gross, Massa, Lerer, Bradbury, Chanan,
  Killeen, Lin, Gimelshein, Antiga et~al.}]{paszke2019pytorch}
Adam Paszke, Sam Gross, Francisco Massa, Adam Lerer, James Bradbury, Gregory
  Chanan, Trevor Killeen, Zeming Lin, Natalia Gimelshein, Luca Antiga, et~al.
  2019.
\newblock \href {https://pytorch.org/} {Pytorch: An imperative style,
  high-performance deep learning library}.
\newblock In \emph{Advances in Neural Information Processing Systems}, pages
  8026--8037.

\bibitem[{Pimentel et~al.(2019)Pimentel, McCarthy, Blasi, Roark, and
  Cotterell}]{pimentel2019meaning}
Tiago Pimentel, Arya~D. McCarthy, Damian Blasi, Brian Roark, and Ryan
  Cotterell. 2019.
\newblock \href {https://www.aclweb.org/anthology} {Meaning to form: Measuring
  systematicity as information}.
\newblock In \emph{Proceedings of the 57th Annual Meeting of the Association
  for Computational Linguistics}, Florence, Italy. Association for
  Computational Linguistics.

\bibitem[{Pimentel et~al.(2021)Pimentel, Roark, Wichmann, Cotterell, and
  Blasi}]{pimentel-etal-2021-finding}
Tiago Pimentel, Brian Roark, S{\o}ren Wichmann, Ryan Cotterell, and Dami\'{a}n
  Blasi. 2021.
\newblock \href {https://arxiv.org/abs/2104.06325} {Finding concept-specific
  biases in form--meaning associations}.
\newblock In \emph{Proceedings of the 2021 Conference of the North {A}merican
  Chapter of the Association for Computational Linguistics: Human Language
  Technologies, Volume 1 (Long and Short Papers)}, Virtual. Association for
  Computational Linguistics.

\bibitem[{Pimentel et~al.(2020)Pimentel, Valvoda, Hall~Maudslay, Zmigrod,
  Williams, and Cotterell}]{pimentel-etal-2020-information}
Tiago Pimentel, Josef Valvoda, Rowan Hall~Maudslay, Ran Zmigrod, Adina
  Williams, and Ryan Cotterell. 2020.
\newblock \href {https://doi.org/10.18653/v1/2020.acl-main.420}
  {Information-theoretic probing for linguistic structure}.
\newblock In \emph{Proceedings of the 58th Annual Meeting of the Association
  for Computational Linguistics}, pages 4609--4622, Online. Association for
  Computational Linguistics.

\bibitem[{Rabelo et~al.(2020)Rabelo, Kim, Goebel, Yoshioka, Kano, and
  Satoh}]{coliee}
Juliano Rabelo, Mi-Young Kim, Randy Goebel, Masaharu Yoshioka, Yoshinobu Kano,
  and Ken Satoh. 2020.
\newblock \href {https://link.springer.com/chapter/10.1007/978-3-030-58790-1_3}
  {A summary of the {COLIEE} 2019 competition}.
\newblock In \emph{New Frontiers in Artificial Intelligence}, pages 34--49,
  Cham. Springer International Publishing.

\bibitem[{Rogers et~al.(2020)Rogers, Kovaleva, and
  Rumshisky}]{rogers2020primer}
Anna Rogers, Olga Kovaleva, and Anna Rumshisky. 2020.
\newblock \href {https://doi.org/10.1162/tacl_a_00349} {A primer in
  {BERT}ology: What we know about how {BERT} works}.
\newblock \emph{Transactions of the Association for Computational Linguistics},
  8:842--866.

\bibitem[{Schindler(1962)}]{schindler1962european}
Dietrich Schindler. 1962.
\newblock \href {https://core.ac.uk/download/pdf/233175113.pdf} {The {E}uropean
  convention on human rights in practice}.
\newblock \emph{Washington University Law Review}, page 152.

\bibitem[{Shannon and Weaver(1962)}]{shannon1948mathematical}
C.~E. Shannon and W.~Weaver. 1962.
\newblock \href {https://books.google.co.uk/books?id=IZ77BwAAQBAJ} {\emph{The
  Mathematical Theory of Communication}}.
\newblock University of Illinois Press.

\bibitem[{Sim et~al.(2016)Sim, Routledge, and Smith}]{sim-etal-2016-friends}
Yanchuan Sim, Bryan Routledge, and Noah~A. Smith. 2016.
\newblock \href {https://doi.org/10.18653/v1/D16-1178} {{F}riends with motives:
  Using text to infer influence on {SCOTUS}}.
\newblock In \emph{Proceedings of the 2016 Conference on Empirical Methods in
  Natural Language Processing}, pages 1724--1733, Austin, Texas. Association
  for Computational Linguistics.

\bibitem[{Spriggs and Hansford(2001)}]{overruling}
James~F. Spriggs and Thomas~G. Hansford. 2001.
\newblock \href {http://www.jstor.org/stable/2691808} {Explaining the
  overruling of u.s. supreme court precedent}.
\newblock \emph{The Journal of Politics}, 63(4):1091--1111.

\bibitem[{Theil(1970)}]{theil1970}
Henri Theil. 1970.
\newblock \href
  {https://www.jstor.org/stable/2775440?seq=1#metadata_info_tab_contents} {On
  the estimation of relationships involving qualitative variables}.
\newblock \emph{American Journal of Sociology}, 76(1):103--154.

\bibitem[{Valvoda et~al.(2018)Valvoda, Ray, and Satoh}]{Valvoda18}
Josef Valvoda, Oliver Ray, and Ken Satoh. 2018.
\newblock \href {https://doi.org/10.3233/978-1-61499-935-5-141} {Using
  agreement statements to identify majority opinion in {UKHL} case law}.
\newblock In \emph{Legal Knowledge and Information Systems - {JURIX} 2018: The
  Thirty-first Annual Conference, Groningen, The Netherlands, 12-14 December
  2018}, volume 313 of \emph{Frontiers in Artificial Intelligence and
  Applications}, pages 141--150. {IOS} Press.

\bibitem[{Vaswani et~al.(2017)Vaswani, Shazeer, Parmar, Uszkoreit, Jones,
  Gomez, Kaiser, and Polosukhin}]{vaswani2017attention}
Ashish Vaswani, Noam Shazeer, Niki Parmar, Jakob Uszkoreit, Llion Jones,
  Aidan~N. Gomez, {\L}ukasz Kaiser, and Illia Polosukhin. 2017.
\newblock \href
  {https://papers.nips.cc/paper/2017/hash/3f5ee243547dee91fbd053c1c4a845aa-Abstract.html}
  {Attention is all you need}.
\newblock In \emph{Advances in Neural Information Processing Systems}, pages
  5998--6008.

\bibitem[{Williams et~al.(2020)Williams, Pimentel, Blix, McCarthy, Chodroff,
  and Cotterell}]{williams-etal-2020-predicting}
Adina Williams, Tiago Pimentel, Hagen Blix, Arya~D. McCarthy, Eleanor Chodroff,
  and Ryan Cotterell. 2020.
\newblock \href {https://doi.org/10.18653/v1/2020.acl-main.597} {Predicting
  declension class from form and meaning}.
\newblock In \emph{Proceedings of the 58th Annual Meeting of the Association
  for Computational Linguistics}, pages 6682--6695, Online. Association for
  Computational Linguistics.

\bibitem[{Wolf et~al.(2020)Wolf, Debut, Sanh, Chaumond, Delangue, Moi, Cistac,
  Rault, Louf, Funtowicz, Davison, Shleifer, von Platen, Ma, Jernite, Plu, Xu,
  Le~Scao, Gugger, Drame, Lhoest, and Rush}]{Wolf2019HuggingFacesTS}
Thomas Wolf, Lysandre Debut, Victor Sanh, Julien Chaumond, Clement Delangue,
  Anthony Moi, Pierric Cistac, Tim Rault, Remi Louf, Morgan Funtowicz, Joe
  Davison, Sam Shleifer, Patrick von Platen, Clara Ma, Yacine Jernite, Julien
  Plu, Canwen Xu, Teven Le~Scao, Sylvain Gugger, Mariama Drame, Quentin Lhoest,
  and Alexander Rush. 2020.
\newblock \href {https://doi.org/10.18653/v1/2020.emnlp-demos.6} {Transformers:
  State-of-the-art natural language processing}.
\newblock In \emph{Proceedings of the 2020 Conference on Empirical Methods in
  Natural Language Processing: System Demonstrations}, pages 38--45, Online.
  Association for Computational Linguistics.

\bibitem[{Xu et~al.(2020)Xu, Wang, Chen, Pan, Wang, and
  Zhao}]{xu2020distinguish}
Nuo Xu, Pinghui Wang, Long Chen, Li~Pan, Xiaoyan Wang, and Junzhou Zhao. 2020.
\newblock \href {https://arxiv.org/abs/2004.02557} {Distinguish confusing law
  articles for legal judgment prediction}.
\newblock \emph{arXiv preprint arXiv:2004.02557}.

\bibitem[{Zhong et~al.(2018)Zhong, Guo, Tu, Xiao, Liu, and
  Sun}]{zhong-etal-2018-legal}
Haoxi Zhong, Zhipeng Guo, Cunchao Tu, Chaojun Xiao, Zhiyuan Liu, and Maosong
  Sun. 2018.
\newblock \href {https://doi.org/10.18653/v1/D18-1390} {Legal judgment
  prediction via topological learning}.
\newblock In \emph{Proceedings of the 2018 Conference on Empirical Methods in
  Natural Language Processing}, pages 3540--3549, Brussels, Belgium.
  Association for Computational Linguistics.

\bibitem[{Zupancic(2016)}]{zupancic}
Bostjan Zupancic. 2016.
\newblock \href
  {https://www.gresham.ac.uk/lecture/transcript/download/in-the-context-of-the-common-law-the-european-court-of-human-rights-in-strasbourg/}
  {{In the Context of the Common Law: The European Court of Human Rights in
  Strasbourg Transcript}}.

\end{thebibliography}

\appendix   

\onecolumn
\section{Glossary:}

\begin{table}[h!]
  \centering
  \begin{tabular}{p{0.20\linewidth} p{0.7\linewidth}}
    \toprule
    \textbf{Legal Terms} \\
    \midrule
    \emph{Facts} & The description of what had happened to the claimant. This includes more general description of who they are, circumstances of the perceived violation of their rights and the proceedings in domestic courts before their appeal to ECtHR.\\
    \midrule
    \emph{Arguments} & The judges explanation of why did they decide the case the way they did. This includes citations of previous cases, application of any relevant legal test, development of a new legal test, analysis of the facts etc.\\
    \midrule
    \emph{Precedent} & Cases that have been cited by the judges as part of their arguments. \\
    \midrule
    \emph{\begin{tabular}{@{}l@{}}Ratio Decidendi\end{tabular}} & The reasons for the decision in a case that is binding on the subsequent cases. Also known as the \emph{\emph{ratio}}. What exactly is ratio is contested by legal scholars.\\
    \midrule
    \emph{Obiter Dicta} & The non-binding discussions in the case. Whatever is not ratio. \\
    \midrule
    \emph{Binding} & Judges are expected to adhere to the binding rules of law and decide future access accordingly. \\
    \midrule
    \emph{\begin{tabular}{@{}l@{}} Stare Decisis\end{tabular}} & New cases with the same facts to the already decided case should lead to the same outcome. This is the doctrine of precedent by which judges can create law. \\
    \midrule
    \emph{Caselaw} & Transcripts of the court proceedings. \\
    \midrule
    \emph{ECHR} & European Convention of Human Rights, comprises of the Convention and the Protocols to the convention. The Protocols are the additions and amendments to the Convention introduced after the signing of the original Convention.\\
    \midrule
    \emph{ECtHR} & European Court of Human Rights, adjudicates ECHR cases. \\
    \bottomrule
  \end{tabular}
  \label{glossary:1}
\end{table}

\begin{table*}
  \centering
  \resizebox{350pt}{!}{%
  \begin{tabular}{p{0.25\linewidth} p{0.65\linewidth}}
  \toprule
    \textbf{Selected ECHR Articles} & \\
    \midrule
    Article 2: \\ \emph{Right to life} & Everyone’s right to life shall be protected by law. No one
        shall be deprived of his life intentionally save in the execution of
        a sentence of a court following his conviction of a crime for which
        this penalty is provided by law.\\
    \midrule
    Article 3: \\ \emph{Prohibition of torture} & No one shall be subjected to torture or to inhuman or degrading
treatment or punishment.\\
    \midrule
    Article 4: \\ \emph{Prohibition of slavery and forced labour} & No one shall be held in slavery or servitude.\\
    \midrule
    Article 8: \\ \emph{Right to respect for private and family life} & Everyone has the right to respect for his private and family
life, his home and his correspondence.\\
    \midrule
    Article 9: \\ \emph{Freedom of thought, conscience and religion} & Everyone has the right to freedom of thought, conscience
and religion; this right includes freedom to change his religion or
belief and freedom, either alone or in community with others and
in public or private, to manifest his religion or belief, in worship,
teaching, practice and observance.\\
    \midrule
    Article 10: \\ \emph{Freedom of expression} & Everyone has the right to freedom of expression. This right
shall include freedom to hold opinions and to receive and impart
information and ideas without interference by public authority
and regardless of frontiers. This Article shall not prevent States
from requiring the licensing of broadcasting, television or cinema
enterprises.\\
    \midrule
        Article 11: \\ \emph{Freedom of assembly and association} & Everyone has the right to freedom of peaceful assembly and
to freedom of association with others, including the right to form
and to join trade unions for the protection of his interests.\\
    \midrule
    Article 12: \\ \emph{Right to marry} & Men and women of marriageable age have the right to marry and
to found a family, according to the national laws governing the
exercise of this right.\\
    \midrule
    Article 13: \\ \emph{Right to an effective remedy} & Everyone whose rights and freedoms as set forth in this Convention
are violated shall have an effective remedy before a national
authority notwithstanding that the violation has been committed
by persons acting in an official capacity.\\
    \midrule
    Article 14: \\ \emph{Prohibition of discrimination} & The enjoyment of the rights and freedoms set forth in this
Convention shall be secured without discrimination on any ground
such as sex, race, colour, language, religion, political or other
opinion, national or social origin, association with a national
minority, property, birth or other status.\\
    \midrule
    Article 18: \\ \emph{Limitation on use of restrictions on rights} & The restrictions permitted under this Convention to the said rights
and freedoms shall not be applied for any purpose other than
those for which they have been prescribed.\\
    \bottomrule
    
  \end{tabular}}
  \label{glossary:2}
\end{table*}

\newpage
\section{Facts \& Arguments Examples}

\begin{table*}[h]
  \centering
  \begin{tabular}{|l|p{13cm}|}
    \hline
    Fact & \emph{The applicants, D.P. and J.C., who are sister and brother, are United Kingdom nationals, born in 1964 and 1967 and living in London and Nottingham, respectively...}\\
    \hline
    Argument & \emph{Article 2 of the Convention provides, in its first sentence: “1. Everyone's right to life shall be protected by law. ...” 46. The applicants complain that the authorities failed to protect the life of their son and were responsible for his death...}\\
    \hline
  \end{tabular}
\end{table*}

\end{document}